\DocumentMetadata{} % this is here to eliminate "Package hyperxmp Error: hyperref must be loaded before hyperxmp." error

\documentclass[manuscript,nonacm]{acm/acmart} 

\usepackage{graphicx} 
\usepackage{enumitem}
\usepackage{multicol} 
\usepackage{color} 
\usepackage{float} 
\floatstyle{boxed} 
\newfloat{Table}{htbp}{lop} 

\AtBeginDocument{%
  \providecommand\BibTeX{{%
    \normalfont B\kern-0.5em{\scshape i\kern-0.25em b}\kern-0.8em\TeX}}}

\begin{document}

\title{Toward Improving Binary Program Comprehension via Embodied Immersion: A Survey}

\author{Dennis Brown}
\email{dgb0028@auburn.edu}
\orcid{0009-0003-9782-1568}
\author{Emily Mulder}
\email{semulder@gmail.com}
\author{Samuel Mulder}
\email{szm0211@auburn.edu}
\orcid{0000-0001-9514-7061}
\affiliation{%
  \institution{Auburn University Department of Computer Science and Software Engineering}
  \streetaddress{1301 Shelby Center}
  \city{Auburn}
  \state{Alabama}
  \country{USA}
  \postcode{36849}
}

\renewcommand{\shortauthors}{Brown, Mulder, and Mulder}

\begin{abstract}
% -------------------
% ABSTRACT IMPORT

Binary program comprehension is critical for many use cases but is difficult, suffering from compounded uncertainty and lack of full automation. We seek methods to improve the effectiveness of the human-machine joint cognitive system performing binary PC. We survey three research areas to perform an indirect cognitive task analysis: cognitive models of the PC process, related elements of cognitive theory, and applicable affordances of virtual reality. Based on common elements in these areas, we identify three overarching themes: enhancing abductive iteration, augmenting working memory, and supporting information organization. These themes spotlight several affordances of VR to exploit in future studies of immersive tools for binary PC.

% -------------------
\end{abstract}

\keywords{binary program comprehension, binary reverse engineering, embodied immersion, immersive analytics, virtual reality, cognitive models, cognitive systems engineering, augmented cognition}

\maketitle

% -------------------
% MAIN CONTENT IMPORT
\section{Introduction}
\label{section:introduction}

Understanding how an existing software program operates is critical to many tasks ranging from maintaining and improving that software to mitigating potential malicious impacts of that software. We call this process \emph{program comprehension (PC)}: comprehending of the purpose and effects of the program. While much of the research in program comprehension primarily focuses on source code written in high-level languages, many of the same concepts and processes can be applied to understanding binary programs that lack the original source code. Comprehending non-trivial binary programs---\emph{binary PC}~\footnote{This paper will use ``binary PC'' to refer to the process of comprehending or understanding, a binary program at the assembly language level. This process is also called ``reverse engineering''  or ``program understanding'' of binary programs in some references, and for the purposes of this work, should be considered synonymous with ``binary PC.''}---is quite difficult. Many tools and approaches have been developed over time to expedite the process, e.g., disassemblers, decompilers, profilers, debuggers, etc., and more ambitious projects employing artificial intelligence~\cite{Hexrays2023}~\cite{Ghidra2023}~\cite{Radare2023}~\cite{David2020}~\cite{Maier2019}. However, successfully understanding a complex binary program still ultimately depends on the expertise of a human in the loop, one with rare knowledge and experience. 

One key to improving a human-centric process is understanding its human element and how the human interacts with the system. Cognitive Systems Engineering (CSE) considers holistically the combination of the operator and the machine---the joint cognitive system---and how to make that joint system more effective by reducing cognitive complexity for the human. In this approach, all work performed by humans is a form of cognitive work~\cite{Hollnagel2005}. The work of binary PC includes many disparate cognitive processes. Notable prior research of the cognitive processes involved in performing binary PC casts it as a sensemaking task~\cite{Bryant2012}, with more nascent research studying binary PC at the level of cognitive modeling~\cite{Votipka2020}~\cite{Mantovani2022}~\cite{Dudenhofer2017} and applying Cognitive Load Theory (CLT)~\cite{Smits2022}. Using the findings from our literature survey, we perform an indirect cognitive task analysis~\cite{Crandall2006} of binary PC to identify its unique cognitive skills and demands.

We assert that if we can reduce the cognitive complexity of binary PC tasks via the employment of immersive interaction,  then the overall outcomes of the joint cognitive system performing binary PC can be improved. Employing embodied cognition is posed as one method of improving cognitive processes~\cite{Wilson2002}, and Virtual Reality (VR) is an accessible means of providing embodied user experiences. In particular, Immersive Analytics (IA)~\cite{Chandler2015}~\cite{Dwyer2018}~\cite{Ens2022} is an emergent and promising area of research that uses innovative display and interaction techniques to aid humans in the understanding of complex analytic problems by providing an egocentric experience. IA may provide the conceptual building blocks, applied in a CSE framework, for significantly improving the task of binary PC.

Toward the improvement of cognitive processes in binary PC using immersive experiences, this work seeks to answer the following Research Questions (RQ):
\begin{enumerate}[label=RQ\arabic*., leftmargin=1.25cm]
    \item What are common and significant characteristics of mental/cognitive models employed by binary PC practitioners?
    \item What are effective techniques for understanding and improving cognition in the context of binary PC? 
    \item How can the affordances of immersive interaction with virtual reality be employed to improve cognition in analytic tasks?
    \item How can we use these findings to effect improvements in the practice of binary PC?    
\end{enumerate}

This work reviews and ties together research contributions from multiple disciplines in order to answer the RQs. These disciplines include Reverse Engineering and Program Comprehension; Cognition and Cognitive Systems; Visualization; VR; and IA. The papers covered in this survey were identified through an iterative approach. We first performed exploratory ad-hoc searches of the literature using common scholarly search engines and indices looking for directly-related results in the primary topic area of using VR to improve binary PC. While we found very few close matches, we did find many adjacent works and through a clustering process, we identified the disciplines listed previously. We then iterated this search process until we filled out a set of recent and relevant research papers. 

This paper proceeds as follows. Section~\ref{section:pchard} dives into what makes binary PC difficult. In Sections~\ref{section:cogmodel} and ~\ref{section:cognition}, we explore prior research in cognitive models of PC (primarily binary and adjacent forms of PC) and the cognitive theory that relates to those models in order to answer RQ1 and RQ2. In Section~\ref{section:immersion}, we explore how affordances in visualization and immersion have been applied toward similar analytic problems to answering RQ3. Finally, we analyze the findings, address RQ4, and consider future work in Section~\ref{section:analysis}.

\section{Why is Binary PC Difficult?}
\label{section:pchard}

Comprehending the actions and purpose of an arbitrary binary program is a complex task. To support our cognitive task analysis---identifying the cognitive skills and mental demands of performing a task~\cite{Militello1998}---and before we dive into the cognitive models and related theory in upcoming sections, we should identify \emph{why} binary PC is inherently difficult. This analysis supports the first of three principles of CSE per Hollnagel and Woods, which is to identify the problem areas of the task. ~\cite{Hollnagel2005}.  

First, this task is very unlikely to be fully automated. Rice's Theorem~\cite{Rice1954} implies that deciding whether a given binary program contains any non-trivial property is formally undecidable.  In practice, with intermingled code and data on modern architectures, there are many common code patterns that are unresolvable. Even with state-of-the-art tools (including Large Language Models), the process of disassembly introduces uncertainty and error.  Each additional property built on top of disassembly is thus both undecidable and built on a representation including that uncertainty. Human input is required to overcome ambiguous situations and determine when a solution is good enough to satisfy the current objective, and even with human input our ability to answer questions about binary programs is severely limited.

Second, understanding what properties we care about in a program is hard; e.g., how do we know if there is a hidden trigger or a backdoor? Even given a source representation, these questions are challenging to answer, often reducing to a question of program correctness.  Since programs rarely have a detailed formal specification, the \emph{correct} behavior may be an open question. Reverse engineering a piece of software is an iterative, hypothesis-driven process, requiring the analyst to continually reformulate and ask questions relative to the current state of knowledge.
% compilation is a one-way transformation

Finally, programs are arguably the most complex things ever engineered by humans~\cite{Crockford2008}. In building complex software, the field of software engineering has developed many models for managing abstraction and interfaces between different components to allow many programmers to design a program as a team.  Without that infrastructure in place, an entire non-trivial program is too complex for a human mind to fully understand. This is compounded by the fact that the compilation process strips helpful comments, structures, and metadata, leaving the analyst to try and recreate the layers of abstraction that the original authors used to manage the complexity.

Next, we will review mental/cognitive models used by analysts to overcome these difficulties when performing binary PC.

\section{Characteristics of Mental/Cognitive Models of Binary PC}
\label{section:cogmodel}

In this section and the next, we examine cognitive elements of binary PC in two phases. In this first phase, we present findings regarding the cognitive process of binary PC. Binary PC shares many concepts and theory with source code PC, which has been more extensively covered in the literature, so we will include some applicable work in source code PC in this section. These sections support the second of the three principles of CSE, which is to understand the circumstances of the problems identified per the first principle, focusing on potential causes and influences of those problems.~\cite{Hollnagel2005} 

\subsection{Discovering Mental/Cognitive Models of Binary PC}

The \emph{mental model}, or cognitive model, concept originated with Craik~\cite{Craik1943}, who asserted that humans build internal representations of the external world and use them to reason about why things happen and anticipate what might happen next. Johnson-Laird's~\cite{JohnsonLaird1983} research in experimental psychology corroborated Craik's claims by finding that people employ mental models in their working memory to perform reasoning.~\cite{Preece2019} Cognitive models can be particularly powerful because they combine elements of both the problem domain or task area and the characteristics of human cognition.~\cite{Anderson1998}

\begin{figure}[htbp]
    \centerline{\includegraphics[width=4.5in]{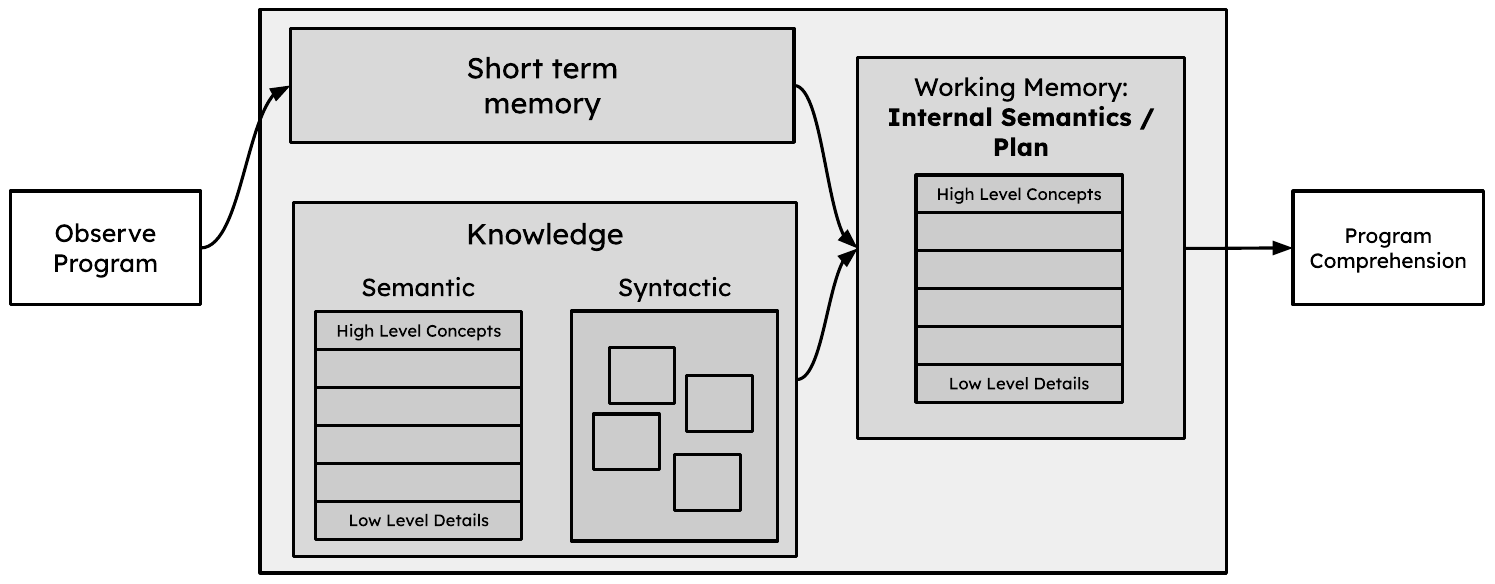}}
%    \centerline{\includesvg[width=4.5in]{images/Shneiderman_Mayer_PC_diagram_2.svg}}
    \caption{PC process per Shneiderman and Mayer~\cite{Shneiderman1979}} 
    \label{fig:pc-shneiderman}
\end{figure}

Shneiderman and Mayer~\cite{Shneiderman1979} developed an early cognitive model of how programmers build up an internal semantic representation of a program based on evidence discovered in a series of experiments they performed. They asserted that the model must describe common programming tasks (composition, comprehension, debugging, modification, and learning) in terms of cognitive structures that programmers form in their memories and cognitive processes to use and build that knowledge. Knowledge is categorized as semantic (general concepts independent of language) or syntactic (``precise, detailed, and arbitrary'' details, primarily language-specific). They cast the task of PC as subtasks of debugging, modification, and learning in which the programmer uses syntactic knowledge to form a multi-level internal semantic representation of the program. Lower levels (e.g., sequences of operations) and higher levels (what the program does) can be formed independently, and this encoding process is similar to Miller's chunking process~\cite{Miller1957}, where smaller chunks of statements join to form larger chunks. The internal semantic representation of the program is strongly retained and widely accessible. Figure~\ref{fig:pc-shneiderman} illustrates their concept. 

Introducing an iterative model, Brooks~\cite{Brooks1983} envisioned the PC process as forming a mental tree that branches into increasingly complete hypotheses of how the code works and what it does. As a branch is found to be incorrect, the programmer cuts that branch and backtracks and may follow another higher-level branch or add a new one. This process is essentially abductive reasoning applied to code understanding, described by Weigand and Hartung~\cite{Weigand2012} as a process of making observations, forming hypotheses, creating mental models of code that support the hypotheses, and searching for information to prove or disprove the hypotheses. 

Zayour and Lethbridge~\cite{Zayour2000} performed a cognitive analysis to identify the mental steps of source-code-based PC. In their analysis, the language was low-level proprietary code and not a typical high-level language on which PC is performed, increasing the applicability of their results to binary PC. They identified two primary cognitive difficulties of disorientation following recursions and disorientation in understanding the most relevant execution paths in the code. In response, they proposed two high level cognitive design requirements. The first is to minimize how many artifacts the engineer needs to keep in working memory by maintaining visual proximity between artifacts; linking new and existing information with meaningful encoding or chunking; and facilitating backtracking in execution paths. The second is to minimize fading of working memory by reducing the time artifacts need to be maintained in working memory and minimizing the number of steps between artifact acquisition. They used their findings to drive implementation of a PC assistive process in their DynaSee tool, which performs several filtering steps on program execution (remove redundancies, detect patterns, rank code routines) before visualizing trace patterns.

\begin{figure}[htbp]
    \centerline{\includegraphics[width=5in]{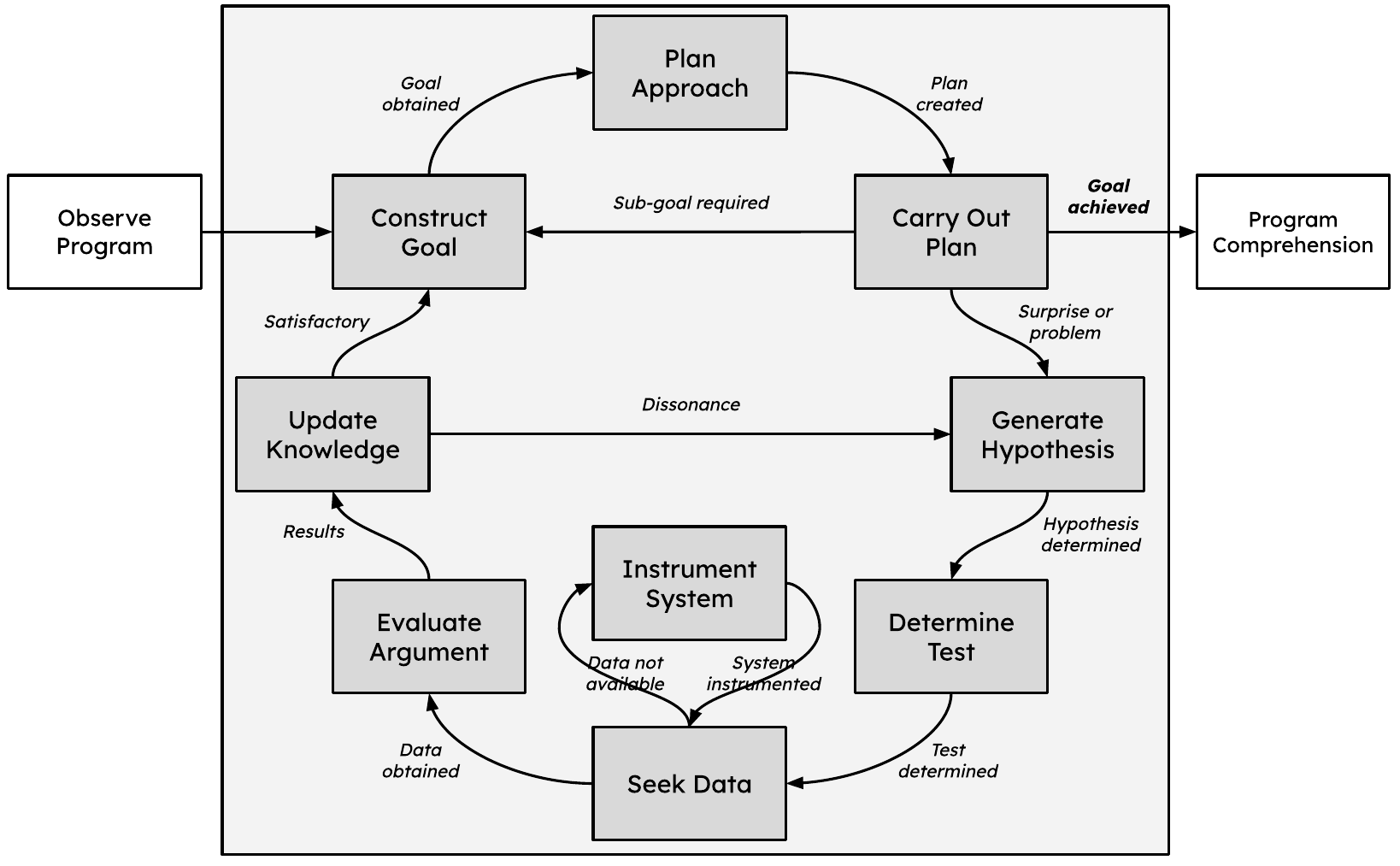}}
    % \centerline{\includesvg[width=5in]{images/Bryant_PC_sensemaking_diagram_2.svg}}
    \caption{PC process based on sensemaking; adapted from Bryant et al.~\cite{Bryant2012}}
    \label{fig:re-bryant-sensemaking-functions}
\end{figure}

The Data-Frame Theory was established by Klein et al.~\cite{Klein2007} in which a frame---an explanatory structure such as a story, map, script, or plan that relates entities to other entities---is simultaneously fitted to discovered data and also drives the discovery of further data. The frame provides a foundation for understanding until flaws are detected.  \emph{Sensemaking} happens when the frame is re-fitted to new evidence in response to those flaws, and it is enabled by abductive reasoning~\cite{Dudenhofer2019}.The approach to modeling taken by Bryant et al.~\cite{Bryant2012} is to frame binary PC as a sensemaking task, in which the engineer develops a mental model or hypothesis about the situation and its constituent elements, and adapts that model through iterations of goal-directed information seeking.  Through a process of cognitive task analysis and verbal protocol analysis, they identified nine sensemaking functions in binary PC and developed a state machine representing the process, represented in Figure~\ref{fig:re-bryant-sensemaking-functions}. Within their model, Bryant et al. included both procedural and declarative knowledge. Procedural knowledge consists of stored patterns of interaction. They categorized the declarative, or factual, knowledge into twelve subdomains covering the fundamental training and experience needed by reverse engineers (programming, debugging, program loading and execution, instruction sets, etc.). Additional declarative knowledge comes from abstract and concrete causal relationships within the data, e.g., constraints on prior knowledge or predicates applied to constants. They explored codifying declarative and procedural knowledge as an implied state machine via production rules implemented in the ACT-R~\cite{Ritter2019} cognitive architecture.

To facilitate and encourage collaboration through communication of information in analysts' mental models, Tennor~\cite{Tennor2015} surveyed and analyzed cognitive aspects of binary PC and proposed that the binary reverse engineering community build vocabularies and ontologies. In this vein, Sisco et al.~\cite{Sisco2017} sought to mathematically formalize Bryant et al.'s concept of binary PC as a sensemaking task and developed a supporting ontology. They asserted that reverse engineers analyze programs using four foundational patterns: navigation (searching for items, e.g., beacons~\footnote{The \emph{beacon} is one important concept in the cognitive process identified by Brooks, which is a ``set of features that typically indicate the occurrence of certain structures or operations within the code.''~\cite{Brooks1983}. An example of a beacon is a block of code that interchanges values within an array inside of a loop, which strongly indicates a sorting function. Programmers seek beacons to confirm their hypotheses in the abductive reasoning process.}); translation (determining how the code would be implemented in a higher-level language); experimentation (deducing how program values change over different flows); and elaboration (identifying and explaining the major components and properties of the program). Reverse engineers use these patterns to build knowledge as interrelated mental objects with various constraints on their relationships. They proposed an ontology for representing this knowledge composed of \emph{ologs}: category-based types, aspects, and facts; this representation allows commutation and a higher level of expressiveness than is supported by common alternatives such as Web Ontology Language (OWL) and Resource Description Framework (RDF). The team generated ologs for fundamental assembly instructions, program data, control flow, and operating system events, and used those ologs to formalize information flow in the experimentation pattern. The team theorized that this work provides the basis for a computational framework to model cognitive tasks in binary PC, and that model can further inform the actions of an analyst or agent encoded in ACT-R.

\begin{figure*}[htbp]
    \centerline{\includegraphics[width=6in]{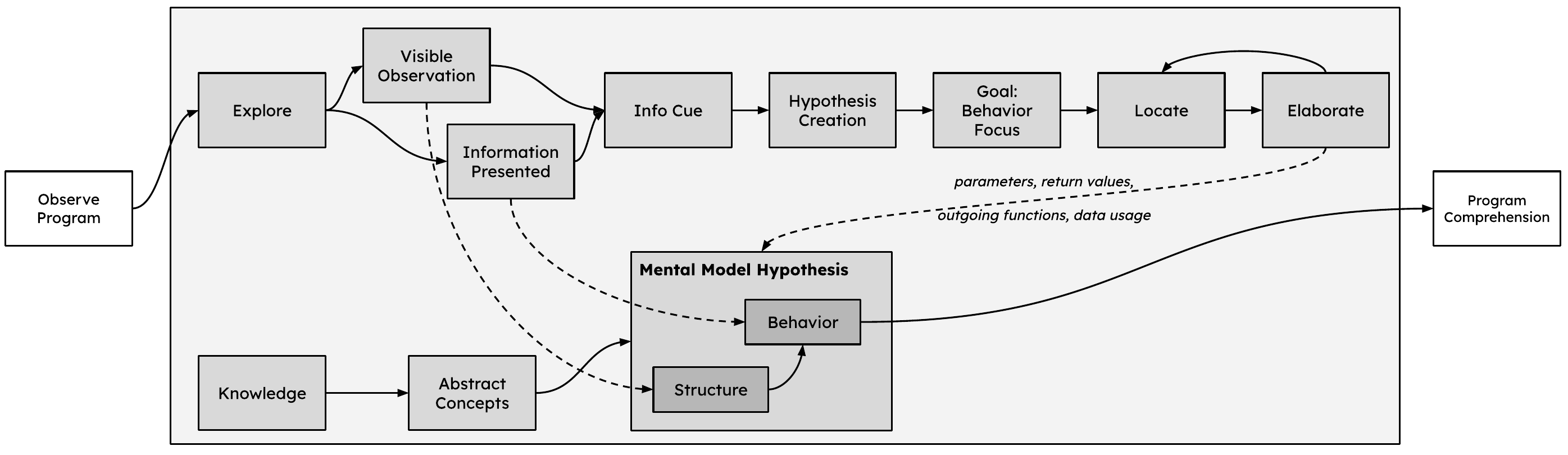}}
    % \centerline{\includesvg[width=6in]{images/Dudenhofer_CURE_diagram_2.svg}}
    \caption{PC process based on abductive iteration; adapted from Dudenhofer~\cite{Dudenhofer2019}}
    \label{fig:re-dudenhofer-cure}
\end{figure*}

In a push toward building automated agents to assist with binary PC, Dudenhofer~\cite{Dudenhofer2019} proposed the Cognitive Understanding of Reverse Engineering (CURE) model to capture sensemaking steps in binary PC. The model, shown in Figure~\ref{fig:re-dudenhofer-cure}, 
is founded on the iterative cycle of abductive reasoning and experimentation: form a hypothesis; explore to find information to prove or disprove a hypothesis; recognize cues in the code or related artifacts, e.g., beacons, and store them in working memory; use that information to refine the hypothesis or form the next hypothesis; and repeat the cycle. The model inspired an application, the CURE Assistant, which works with an existing binary reverse engineering framework to identify code snippets matching those in an extensible catalog of recipes and present possible program behaviors to the analyst. 

\begin{figure*}[htbp]
    \centerline{\includegraphics[width=6in]{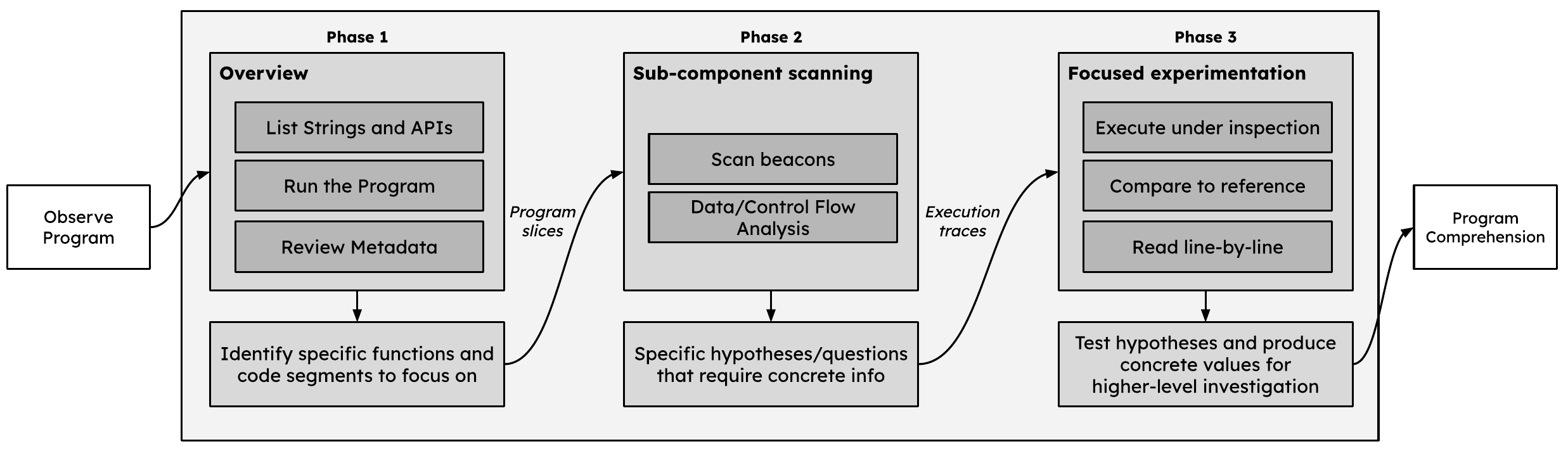}}
    % \centerline{\includesvg[width=6in]{images/Votipka_RE_phases_2.svg}}
    \caption{PC process in three phases, adapted from Votipka et al.~\cite{Votipka2020}}
    \label{fig:re-votipka}
\end{figure*}

Votipka et al.~\cite{Votipka2020} presented a three-phase process model for binary PC: (1) Examining and executing the full program for an overview, then choosing focus areas; (2) Reviewing specific program slices chosen in the first phase, scanning for beacons and data and control flows, and generating specific hypotheses; and (3) Inspecting lines of assembly code and traces to test the hypotheses, as shown in Figure~\ref{fig:re-votipka}. They also identified several categories of beacons for use across the phases: APIs and strings across all phases; UI elements in the first phase; and constants, variable names, control flow structures, compiler optimizations, function prototypes, and program flow in the second and third phases. In developing this process, they identified one significant similarity between binary PC and source-code-based PC: mental simulation of code execution; and two primary differences between binary PC and source-code-based PC: binary PC involves more overview while source-code-based PC is more focused and pinpointed, and binary PC uses a more diverse set of beacons, common recognizable schema or patterns. Their work culminated in five guidelines for designing RE tools: (1) Match interaction with analysis phases; (2) Present input and output in the context of code; (3) Allow data transfer between static and dynamic contexts; (4) Allow selection of analysis methods; and (5) Support readability improvements. 

Nyre-Yu et al.~\cite{Nyre-Yu2022} conducted a task analysis of static binary PC. They found evidence reinforcing previous findings that binary PC and source-code-based PC share similarities in cognitive processes: identifying goals and plans, creating hypotheses, and exploring to gather information. They asserted that analysts very commonly ask ``where is this used in the code?'', ``where is the method being called?'', ``how can I get calling information?'', ``where does this information/data go?'', ``where is the data coming from?'' and ``what is the context of this vulnerability/code?''~\cite{Nyre-Yu2022}. They also observed the revisiting of past states reported by Votipka et al.~\cite{Votipka2020}, however, they did not find common patterns across participants. The team also identified repetitive actions that are targets for automation: accessing library function documentation and accessing the task definition. 

\subsection{Differences between Novices and Experts in Performing Binary PC}

Besides the mental models associated with binary PC, it is important to understand differences between novices and experts. While both classes of analysts use many of the same elements of the mental model, the approaches they take are different. 

Looking toward source-code-based PC again as a close analog of binary PC, Vessey~\cite{Vessey1985} studied experts and novices in a code comprehension and debugging task, observing that while both experts and novices employ breadth-first approaches, experts apply a system view that novices do not; novices also employ depth-first approaches while experts do not. Additionally, experts perform chunking effectively to proceed steadily through a program, while novices perform chunking much less effectively, resulting in jumping around within a program to understand it. Storey's survey~\cite{Storey2005} found that experts use more external memory aids~\cite{Detienne2001}, and that novices focus mainly on objects while experts also consider functional relationships and algorithms. Siegmund et al.~\cite{Siegmund2014} asserted there are two basic models of PC: top-down and bottom-up. Rather than distinguishing between ``expert'' and ``novice,'' they focused on level of domain knowledge. Analysts who have domain knowledge they can apply to the program use the top-down process and use beacons to form hypotheses. Otherwise, without domain knowledge, they use the bottom-up process to analyze the program line-by-line.

Cowley~\cite{Cowley2014} conducted a job analysis to identify performance predictors for binary PC practitioners. In addition to situational (team and organizational) predictors, Cowley identified individual predictors for novices and experts. From those predictors, the author determined five milestones along the progression from novice to expert. Milestones to the intermediate level include (1) proficiency in relevant tools; (2) significant reduction in assistance needed to complete tasks; and (3) parity with experts in identification and employment of binary PC strategies. The final two milestones complete the transition to a true expert, including (4) organizational recognition through promotion; and (5) establishing a track record of solving binary PC problems without assistance. 

Mantovani et al.~\cite{Mantovani2022} studied nearly 300 hours of activity performed by 72 novice and expert reverse engineers performing a static analysis task. They observed that most often, novices move forward from \texttt{main()} and jump around between code blocks, visiting some more than once, while experts move both forward and backward from \texttt{main()} and move more linearly through code blocks. Additionally, experts are quicker to identify what they can ignore.

\subsection{Section Summary}

In this section, we surveyed evidence in support of answering RQ1, ``What are common and significant characteristics of mental/cognitive models employed by binary PC practitioners?'' We examined a progression of research contributions in this area, starting with the basic conceptualization of a mental or cognitive model. Early research in mental models for software development proposed an internal semantic representation and chunking process, then progressed to include iterative hypotheses with abductive reasoning. Researchers then formalized the sensemaking process of goal-oriented iteration and adaptation of the analyst's emergent mental model, and further developed the field by creating an ontology and introducing automation. 

This progression of research yielded a set of common concepts. While researchers found that the process of binary PC is crafted uniquely by each individual, especially at the expert level, we identify the following themes and salient points regarding modeling the cognitive activities involved in binary PC:
\begin{itemize}
    \item Uses multi-level internal semantic representation of the program~\cite{Shneiderman1979}
    \item Follows iterative pattern of sensemaking / abductive reasoning
        \begin{itemize}
            \item Sets goals and follows plans~\cite{Bryant2012}~\cite{Nyre-Yu2022} 
            \item Forms increasingly complete hypotheses~\cite{Brooks1983}~\cite{Weigand2012}~\cite{Dudenhofer2019}~\cite{Votipka2020}
            \item Tests hypothesis through experimentation~\cite{Bryant2012} ~\cite{Sisco2017}~\cite{Dudenhofer2019}~\cite{Votipka2020}
            \item Updates framing of the problem based on results~\cite{Klein2007}~\cite{Bryant2012}~\cite{Dudenhofer2019}~\cite{Votipka2020}
        \end{itemize}
    \item Creates disorientation following recursions and execution paths~\cite{Zayour2000}
    \item Taxes working memory~\cite{Shneiderman1979}~\cite{Zayour2000}
    \item Requires declarative and procedural knowledge retrieval and generation~\cite{Bryant2012}  
    \item Uses translation--determining how the code would be implemented in a higher-level language~\cite{Sisco2017}
    \item Uses beacons; beacons for binary PC are more diverse than for source-code-based PC~\cite{Brooks1983}~\cite{Dudenhofer2019}~\cite{Votipka2020}
    \item Relies upon overview of the binary executable~\cite{Votipka2020}
    \item Uses external memory aids~\cite{Detienne2001}~\cite{Storey2005}
    \item Relies upon determining what to ignore~\cite{Mantovani2022}
    \item Experts use a breadth-first approach with system thinking; novices, both depth- and breadth-first without system thinking~\cite{Vessey1985}
    \item Experts use a top-down approach if the program domain is familiar, otherwise, bottom-up~\cite{Siegmund2014}
\end{itemize}

Continuing our effort to understand the circumstances of the problems of binary PC, we next look at findings in the literature about underlying cognitive theory with the aim of augmenting or improving the cognitive activities of binary PC.

\section{Underlying Cognitive Theory and Applications to Binary Program Comprehension}
\label{section:cognition}

Now that we have a baseline understanding of the characteristics of mental/cognitive models behind binary PC, we explore possible avenues for enhancing the abilities of practitioners. In this section we briefly review the underlying theory behind external and embodied cognition along with cognitive load, and then consider methods that may reduce cognitive load or otherwise optimize the cognitive processes of binary PC practitioners.

\subsection{External and Embodied Cognition}

Cognition involves many specific and interdependent processes: Eysenck and Brysbaert~\cite{Eysenck2018} identified attention; perception; memory; learning; reading, speaking, and listening; and problem-solving, planning, reasoning, and decision-making; each with their own design implications. Norman described cognition as multimodal---that there are many different types of thinking---and specifically addressed experiential cognition and reflective cognition~\cite{Norman1993}. Experiential cognition is when a person is experiencing and responding to the environment without the need for significant mental effort (which also happens with extensive expertise), while reflective cognition happens when a person is putting substantial thought into considering and making decisions or forming new ideas. Norman further discussed both forms in relation to effective tools for enhancing one or the other, such that a tool designed to aid reflective thought is inappropriate for experiential cognition and vice versa~\cite{Norman1993}. Cognition also occurs in varied places, times, and situations, and there is movement in the field toward understanding cognition \emph{in situ}---rather than limiting the scope of reasoning about cognition to what occurs in the mind, considering how the environment can improve and affect cognition~\cite{Preece2019}.

External cognition, per Scaife and Rogers~\cite{Scaife1996}, concerns the cognitive interaction with external manifestations of knowledge in the environment, e.g., images, video, virtual reality, etc. The primary cognitive benefits of external cognitive activities include using external knowledge representations (e.g., notes and reminders) to reduce memory load; using computational tools (e.g., calculators) to make tasks easier; and annotating and reordering or restructuring external representations of knowledge (e.g., checking off to-do lists or arranging desktop icons)~\cite{Preece2019}. In considering the impact of the environment on cognition, Gibson coined the term \emph{affordance} as a ``specific combination of the properties of its substance and its surfaces’’ perceived in relation to the viewer~\cite{Gibson1977}. Norman further elaborated that an affordance is most importantly constrained by properties that indicate how something could be used: ``When affordances are taken advantage of, the user knows what to do just by looking: no picture, label, or instruction needed.’’~\cite{Norman1988}. Affordances leverage internal cognition to enable the phenomena of external cognition and embodied cognition. 

Embodied cognition~\cite{Gelder1996}, closely linked to dynamical systems theory, proposes that cognition happens in real time, with the brain simultaneously receiving input from, processing, and interacting within and with the nervous system, physical body, and external environment. In this theory, cognition is deeply impacted by sensorimotor interaction with the environment and is profoundly grounded in the ability to act~\cite{Glenberg2013}. Shapiro and Spaulding’s \emph{Embodiment Thesis} states: ``Many features of cognition are embodied in that they are deeply dependent upon characteristics of the physical body of an agent, such that the agent’s beyond-the-brain body plays a significant causal role, or a physically constitutive role, in that agent’s cognitive processing.’’~\cite{Shapiro2021}

In line with the concept of the joint cognitive system of human and machine described earlier, Kirsh~\cite{Kirsh2013} made the case that tools are absorbed into the body schema and change how we think, that we think with not just our minds but also our bodies (enabled by kinesthetic perception), and that we even think with tools. Similarly, Hornecker et al.~\cite{Hornecker2017} presented the position that our sensorimotor interactions with, and manipulation of, the world develop our capacity for abstract thought, starting with simple concepts such as in/out, over/under, up/down — that body and mind are inseparable even in the domain of abstract thought. Ale et al.~\cite{Ale2022} suggested that embodied memory is a potentially rich research area, in which physical objects or locations serve as memory palaces; and that whole-body stimuli can expedite storage and retrieval of memory (e.g., using physical motion or aroma triggers to encode and store memories).

\subsection{Cognitive Load Theory}

Cognitive Load Theory (CLT), per Sweller and Chandler~\cite{Sweller1994}, assumes that humans take in and process information through two main channels---through hearing and through visualizing. Further, only a finite amount of processing can occur in each channel at any point in time. The magnitude of the cognitive load depends not only on sheer number, but on how much interactivity occurs between elements. The ability to process this information is limited by working memory, but can be improved through employing a suitable schema with which to leverage long-term memory and reduce cognitive load. Cognitive load theory currently incorporates three categories: \emph{intrinsic} load inherent in the cognitive task at hand; \emph{extraneous} load caused by how information is presented; and \emph{germane} load. Germane load represents the ability of our mind to connect what we are learning with long term memory, and linked to intrinsic, is the demand on our cognition of using or focusing working memory for intrinsic learning~\cite{Sweller2019}.  Paas et al.~\cite{Paas2004} asserted that cognitive performance will degrade at either end of the load spectrum (underloading or overloading).

Hollender et al.~\cite{Hollender2010} surveyed 65 papers on cognitive load related to human-computer interaction. They cataloged methods to leverage phenomena to reduce extraneous cognitive load as follows. The worked example effect is from learning from studying solved sample problems. The split-attention effect is from presenting information from multiple visual sources in an integrated way to reduce load required to perform mental integration. The modality effect is from presenting multiple information sources through different modalities (e.g., visual and aural) to allow the inputs in each modality to be processed simultaneously. The redundancy effect is from reducing the level of redundant information presented in different modalities/sources, thereby reducing the load of reconciling the underlying concepts across the inputs from those modalities. They also reviewed methods to increase germane cognitive load and foster schemata development (most applicable in educational settings, and a way to increase the capacity of working memory): specifically introducing a variety of tasks; linking concrete information to abstract concepts; and self-explanation. Also, in an educational context, they reviewed methods to adjust intrinsic cognitive load via adjusting the sizes and quantity of information chunks presented over time.

\subsection{Applications to Program Comprehension}

Next, we will review contributions from studies applying underlying cognitive theory to PC and closely-related tasks. Helgesson and Runeson~\cite{Helgesson2021} studied cognitive load drivers in software engineering and proposed a set of perspectives with which to reason about these drivers. The \emph{Task} perspective accounts for CLT's intrinsic load, which results from the inherent cognitive intensity of software engineering. The \emph{Environment} perspective accounts for CLT's germane load, which results from constructing mental schemata for processes and tools, plus additional load from re-learning new processes and tools that are meant to replace old ones, but can set up competing mental schemata. The remaining perspectives comprise CLT's extraneous load: \emph{Structural} (e.g., technical debt); \emph{Information} (e.g., poor/missing code documentation); \emph{Tool} (e.g., friction from unintuitive, cumbersome, or unreliable tools), \emph{Communication} (e.g., lack of communications amongst the development team), \emph{Interruption} (the cognitive cost due to resumption lag), and \emph{Temporal} (e.g., tracing a component's change history in version control). 

In observing how programmers comprehend code, Siegmund et al.~\cite{Siegmund2014} took a novel approach by employing functional magnetic resonance imaging (fMRI) to find the brain regions activated during source-code-based PC. In a controlled study of 17 computer science undergraduates, designed to elicit bottom-up source code comprehension and minimize extrinsic cognitive load, they observed activation of five Brodmann areas (of 52 Brodmann areas associated with cognitive process)~\cite{Brodmann2006} using fMRI. Those five areas are associated with division of attention, silent word reading, verbal/numerical working memory, and problem solving. Based on further analysis, Siegmund et al. theorized that bottom-up PC uses two areas for keeping values in mind, another area for analyzing words and symbols, and the remaining two areas for integrating statements and chunks. 

Smits~\cite{Smits2022} performed a study of reducing cognitive load in reverse engineering. Under the assumption that the complexity and volume of outputs from RE tools induces cognitive overload, Smits implemented two primary techniques for managing cognitive overload: (1) \emph{Information Filtering} to remove extraneous information so as to reduce extraneous load; and (2) \emph{Information Organization} to organize data in a manner familiar to the user to reduce germane load. The two techniques would therefore increase the capacity available for intrinsic load. The implementation focused on improving the common Control Flow Graph (CFG) visualization with the concept of the \emph{Proximity View}: Simplify the view by removing most instructions, variables, and constants; keep only variables and constants that are arguments for function call nodes; and insert empty nodes to maintain the graph structure. A user study of 41 participants comparing this view to a traditional view demonstrated that subjects in the Proximity View group had statistically significant better performance in challenges solved, but took longer to solve them.

\subsection{Section Summary}

In this section, prompted by RQ2, ``What are effective techniques for understanding and improving cognition in the context of binary PC?'', we reviewed relevant cognitive theory and its applications to binary PC. Starting with the basic concept of cognition, we progressed through external cognition, embodied cognition, and the concept of affordances, which are especially relevant to understanding how immersive interaction can influence and improve accomplishing analytic tasks. We then reviewed Cognitive Load Theory and some of its prior applications in software engineering and program comprehension. 

In reviewing the prior research, we identified the following primary concepts and techniques that stood out, some across multiple sources, as being especially relevant to binary PC:
\begin{itemize}
    \item External Cognition
    \begin{itemize}
        \item Use external knowledge representations to reduce memory load, e.g., notes and reminders~\cite{Scaife1996}~\cite{Preece2019}
        \item Use computational tools (e.g., calculators) to make tasks easier~\cite{Scaife1996}~\cite{Preece2019}
        \item Annotate and reorder or restructure external representations of knowledge~\cite{Scaife1996}~\cite{Preece2019}
    \end{itemize}
    \item Embodied Cognition and Memory
    \begin{itemize}
        \item Cognitive processing is influenced by the body and sensorimotor interactions~\cite{Glenberg2013}~\cite{Hornecker2017}~\cite{Shapiro2021}
        \item Tools extend the body schema~\cite{Kirsh2013}
        \item Physical objects or locations serve as memory palaces~\cite{Ale2022}
        \item Whole-body stimuli can expedite storage and retrieval of memory~\cite{Ale2022}
    \end{itemize}
    \item Cognitive Load Theory
    \begin{itemize}
        \item Balance immediate problem-solving (intrinsic load) and long-term schema development (germane load)~\cite{Sweller2019}
        \item Worked example effect of learning from studying solved sample problems~\cite{Hollender2010}
        \item Split-attention effect of presenting information from multiple visual sources in an integrated way to reduce load required to perform mental integration~\cite{Hollender2010}
        \item Modality effect of presenting multiple information sources through different modalities (e.g., visual and aural) to allow the inputs in each modality to be processed simultaneously~\cite{Hollender2010}
        \item Remove redundancy of information presented in different modalities/sources to reduce the load of reconciling the underlying concepts across the inputs from those modalities~\cite{Hollender2010}
        \item Both underloading and overloading can degrade performance; underloading is unlikely due to the high intrinsic load of PC~\cite{Paas2004}~\cite{Helgesson2021}
        \item Reduce extraneous load: Use human-centered design to reduce the wasted time and friction of poor tools and interactions used to solve a problem~\cite{Helgesson2021}
    \end{itemize}
    \item PC activates areas of the brain associated with working memory, written language, and integration~\cite{Siegmund2014}
\end{itemize}

Now that we have reviewed elements of cognitive theory that underlie the cognitive processes of binary PC, we will examine how immersive technologies can improve those cognitive processes.

\section{Cognitive Augmentation using Visualization and Immersion}
\label{section:immersion}

The third of three principles of CSE~\cite{Hollnagel2005} directs that we should pursue practical solutions to the problem areas that we identified and further developed in Sections~\ref{section:pchard} through~\ref{section:cognition} by following the first and second principles of CSE. In pursuit of those solutions, we present in this section prior work in applying immersive technologies to improve cognition and performance in analytic tasks in general, with a few examples specifically for program comprehension and understanding. 

\subsection{Leveraging Affordances of Immersive Technologies for Analytic Tasks}

As described in Section~\ref{section:cognition}, affordances are perceivable aspects of the environment
that foster interaction and enable external and embodied cognition. Some common affordances of virtual, augmented, and mixed reality (VR, AR, MR) include immersive visual and aural displays, spatial interaction / motion tracking, gesture and voice recognition, haptic feedback, plus additional emerging modalities. In this subsection, we look at prior work in using the affordances of immersive technologies to improve cognitive processes used by analytic and sensemaking tasks. 

One key feature of an immersive environment is an abundance of space. Although the work of Andrews et al.~\cite{Andrews2010} focused on the use of large two-dimensional displays, they proposed several benefits that working in a large spatial environment brings to the sensemaking process. They completed an observational study of participants performing well-known data analytic tasks using their desktop computing environment, which included a 32-megapixel display of approximately 100 inches by 31 inches. Generally, participants used the space to arrange and organize documents and applications in ways that reflected their relationships. The study showed evidence of a number of avenues to exploit for improving cognition by using a large spatial environment including persistence, context, physical navigation, presence of detail, memory refresh, situational awareness, and spatial semantics. These concepts are listed in full in the section summary, citing the authors, and will be further explored in Section~\ref{section:analysis}.
The findings by Andrews et al. endorse concepts we covered in Section~\ref{section:cognition} regarding external cognition and memory (augmenting working memory and structuring external knowledge representations) and embodied cognition (exploiting and extending the body schema).

In their hybrid survey/position paper, Moloney et al.~\cite{Moloney2018} consider the question of whether VR technology provides affordances for analyzing big data. Their survey covers 47 selected publications in the field of visual data mining, which employs visual cognition to augment algorithmic analysis by differentiating variables by mapping them to distinct graphic attributes to harness human visual perception and creativity. Starting with the Computer Aided Virtual Environment (CAVE) in 1992, they provide a timeline of advances through `VR 2.0' and IA, and present future research challenges proposed by research teams. The authors perform an analysis of affordance theory in immersive VR as it exists in the literature and present a position on the shift from allocentric (attention focused externally) visual analytics---traditional visual analytics~\cite{Cui2019}, which is performed using a 2D screen, even for 3D visualizations---to egocentric (attention focused internally) spatial coding and the affordance of VR. The key principles they identified include tuning the environment to human perception; using mimetic/naturalistic references; coordinating multiple modes of interaction; and aligning data selection with naturally-occurring distribution patterns. 
These principles are listed in full in the section summary, citing the authors, and will be further explored in Section~\ref{section:analysis}.

Gra{\v{c}}anin~\cite{Gracanin2018} makes the case that VR and MR technology can provide significant advantages in forming insights about complex data sets by leveraging the theory of affordances to increase embodied resources. The author proposes a framework for creating stimuli in an MR environment based on data from sensors and the user's interactions with the system, which can reveal the most effective immersive and embodied stimuli through an iterative feedback process. Gra{\v{c}}anin concludes that the immersion provided by VR may not be sufficient to fully leverage embodied cognition, and that the physical/real-world affordances provided by MR resolve that gap. Billinghurst et al.~\cite{Billinghurst2019} similarly have developed a VR system incorporating a feedback loop using sensor data to adapt the interaction. Their work has demonstrated that electroencephalogram (EEG) measurements can assess cognitive load in VR training tasks, and the simulations can be adapted based on those measures to benefit training transfer effects. Ahmadi et al.~\cite{Ahmadi2023}, in related research, further refined the findings to pinpoint that power spectral density measurements of the EEG alpha band may reliably indicate cognitive load in moderate VR tasks. 

Batch et al.~\cite{Batch2020} conducted an experiment to determine the effects of using IA for tasks in economic analysis. They used an iterative, human-centered approach with subject matter experts to extend the ImAxes tool~\cite{Cordeil2017}, which provides embodied data axes in a VR environment. They employed a VR headset instead of AR due to better resolution and field of view. Some of their findings were surprises---e.g., fatigue was not a significant factor; no clear concerns about legibility; and participants with little gaming/VR experience used the tool effectively. Participants created egocentric presentation layouts, but did not appear to use the physical space to group similar views near each other, simply using the closest free space. Although this observation contradicted their hypothesis, the authors assert that what they did observe is consistent with the `sensemaking loop' described by Pirolli et al.~\cite{Pirolli2005}: in the early bottom-up search for information when executing a task, retained information is left somewhat unstructured except to store it in a ``shoebox'' for later processing. These findings of Batch et al. corroborate those of Andrews et al.~\cite{Andrews2010} when employing large two-dimensional displays for sensemaking tasks. Ultimately, Batch et al.~\cite{Batch2020} concluded that participants still did not fully use the three-dimensional space available in the environment beyond their immediate proximity, and speculate causes may include the small size of the physical space in which the study was conducted; ability to perform the techniques or gestures required to move the visualizations; and lack of automatic layout, indicating a need for constraints and organization frameworks. 

Prather et al.~\cite{Prather2020} surveyed 104 papers on cognitive augmentation using immersive technologies, specifically work employing biosensor-based measures of user cognitive capabilities in immersive and semi-immersive environments. The authors performed this survey to support their aim of designing and developing cross-reality (XR) systems where task parameters are adapted to optimize the user's cognitive load, particular in `Industry 4.0' use cases where machines carry out repetitive and increasingly complex tasks. These systems would perform as intelligent cognitive assistants to enhance human capabilities. Their survey found that existing research is predominantly in the health and rehabilitation domain rather than industrial engineering. Many efforts target affective/emotional wellness with an emphasis on short-term therapeutic tools rather than an enduring assistive system. Over one third of papers described work in adapting the user experience based on physiological data (predominantly electroencephalography (EEG)), with a small but significant number of them applying Artificial Intelligence (AI) and Machine Learning (ML) to that data. The authors included no papers that directly addressed CLT, but several addressed mitigation of mental exertion.  

Some recent experimental findings in research on immersive technology and cognitive load pinpoint where the technology is most and least effective. Frederiksen et al.~\cite{Frederiksen2020} performed a study of cognitive load in surgical training using immersive and non-immersive VR. The immersive VR method increased cognitive load significantly more than non-immersive VR in both nonstressor and stressor phases, which was attributed to extraneous cognitive load due to the high level of interaction with immersive VR elements. De Melo~\cite{deMelo2020} et al. sought to reduce cognitive load with a virtual embodied AI-based assistant that used a human-appearing avatar in the virtual world to provide guidance to participants in completing tasks. In a controlled experiment comparing the embodied assistant to voice-only assistant and no assistant, embodied assistants led to lower cognitive load than voice-only assistants and both were lower than no assistant. Albus et al.~\cite{Albus2021} studied the effects of the signaling principle in VR learning environments. Per Mayer~\cite{Mayer2005}, using signals to direct a user to pay attention to specific information can create deeper understanding; this is the signaling principle. Compare this concept to beacons described earlier. In their study of a learning exercise with and without annotations (signals), Albus et al. found that the signals increased germane cognitive load, but did not reduce extraneous cognitive load, and did not result in significantly better deep understanding. Chen et al.~\cite{Chen2022} performed a controlled experiment measuring engineering creativity in a secondary education setting, with and without VR. They found that while VR improved cognition, motivation, and the novelty and usefulness of the designs, it did not improve creative thinking. Additionally, VR improved extraneous and germane cognitive load, but had no effect on intrinsic load. 

The field of IA emerged only recently~\cite{Chandler2015}~\cite{Dwyer2018} with an aim to immerse users in their data to improve sensemaking and formation of insights for real world data analytic tasks. Most recently, Ens et al.~\cite{Ens2022} have proposed `Immersive Analytics 2.0.' They recognize that evidence found in prior work in IA points to benefits resulting from several areas: ``use of unlimited space around the users, spatial memory, proprioception [awareness of body motion], 3D spatial interaction, tangible interaction [props and haptics], collaboration, kinesthesia [body motion], [and] engagement.'' They call for work to more deeply understand the motivations for spatial and embodied interaction, and to clearly compare immersive versus non-immersive experiences in analytics.

\subsection{Visualization Applied to Program Comprehension}

Employing visualization can impact the effectiveness of analysts in binary PC and how they exercise their cognitive models. This section covers prior work in visualization to support PC, including that for binary PC and applicable work in source-level PC.

The bulk of prior work in this application area is aimed at the 2D desktop interaction metaphor, and we provide a few notable examples. Early work by Waguespack~\cite{Waguespack1989} implemented visualizations to aid novice Pascal programmers, for example, representing data types as different shapes, and using a variety of visual representations to differentiate type declarations, variable instances, and literal values. Structures are represented as containers of constituent components. This work demonstrated the utility of \emph{chunking}: collecting lower-level details in a single higher-level abstraction that can help understanding of a large problem, then providing the ability to decompose as necessary for detailed analysis. Conti et al.~\cite{Conti2008} implemented a 2D application to visualize binary and data files as bitmaps; e.g., representing one byte per pixel with shading based on the byte's value or presence in an address range. Rendering a binary file in that way, compared to viewing it in a hex editor, gives a view of the entire file at a glance at the expense of low-level details. The method complements the text-based viewer to quickly identify major segments in the file, recurring or unusual patterns, and so on. Gregio et al.~\cite{Gregio2012} demonstrated two 2D/pseudo-3D visualization methods of behavior of suspected malware: timeline plots and icons arranged in a spiral. Both methods display operating system actions taken by the code on objects (e.g., read file; terminate process; etc.) over a period of time in a single, compact view that complements traditional logs. The visualizations were particularly useful in identifying where two sample programs acted similarly, indicating shared code across different programs, or different revisions of the same program. 

Considering a narrower application area, Wagner et. al.~\cite{Wagner2015} surveyed a pool of 220 papers related to code visualization and identified 25 papers specifically about malware visualization systems. From those papers, they identified nine \emph{data providers} behind those visualization systems. The data providers collect information about the suspected malware and provide it to the visualization systems---automated and manual applications that execute static and dynamic analysis techniques to collect data useful in profiling and classifying the malware. The visualization systems were binned into three broad groups: individual malware analysis, malware comparison, and malware summarization. Additionally, they were categorized based on well-established taxonomies from the visualization community: the type of provided data, visualization techniques used, mapping and representation space, temporal aspects, interactivity, and goal/action. They identified challenges in bridging between the three broad groups, integrating disparate data sources, characterizing and abstracting problems, improving expert interaction, and integrating analytical methods with the visualizations. These systems are still aimed at the 2D desktop interaction metaphor---while some of the systems analyzed employed 3D visualizations, none were reported in this survey to employ virtual reality or immersive (embodied/egocentric) techniques.

Elliott et al.~\cite{Elliott2015} explored the use of VR in software engineering to address problems in navigating and comprehending code. Their work builds upon prior research in how developers use the affordance of \emph{spatial memory} in traditional 2D development environments, such as using scrollbar and tab positions as cues~\cite{Ko2006}, or using an ``infinitely''-scrollable document canvas~\cite{DeLine2010}~\cite{Bragdon2010}. This work extends that concept to the affordances of VR applied to software development: spatial cognition, cues, and presence; manipulation and motion to improve perception and retention; and immediate feedback on the state of the system. With these affordances implemented in their RIFTSKETCH (live development) and IMMERSION (code review) tools, the authors provide a proof-of-concept and a vision of VR-based development in the future, though no formal user studies were conducted. 

In a user study conducted by Dominic et al.~\cite{Dominic2020}, 26 graduate students were tested on comprehending simple Java programs of the type that one may find in first-year programming course homework assignments. They compared the traditional desktop experience with a VR configuration entitled "VirtualDesk" using a headset and tracked keyboard and mouse that were mapped 1:1 to the real world. The study did not implement specific affordances of VR as proposed by Elliott et al.~\cite{Elliott2015}, but instead compared the performance of PC using common 2D tools on a traditional desktop environment against using the same tools in VR. Their results show 75\% of programs were comprehended correctly in the traditional desktop experience compared to 65\% in VR. Additionally, their results from conducting the NASA Task Load Index (TLX) survey~\cite{Hart1986} showed significantly more task load---the demand or difficulty in performing a task---in VR. Finally, results of a survey of self-reported concentration and productivity showed that users in VR had lower levels of concentration and no significant difference in perceived productivity. 

One rich facet of human-computer interface theory is the use of metaphors to influence the design of affordances. Lakoff~\cite{Lakoff1994} curated an extensive list of metaphors encountered in linguistics. Metaphors leverage common experiences amongst most users to facilitate understanding of new concepts. Many of these metaphors can apply to perception and cognition in interactive applications, e.g., ``seeing is touching,'' ``the visual field is a container,'' ``theories are constructed objects,'' etc. Averbukh et al.~\cite{Averbukh2019} surveyed applications of VR for (high level) program visualization and visual programming, and in particular, the metaphors employed in those applications. They reviewed the city, molecule, and heliocentric cosmic metaphors, asserting they share important qualities: ``unlimited context, organization of inner structure, naturalness, and resistance to scaling,'' and that these natural metaphors simplify spatial orientation and navigation in the VR world.  

The city metaphor recurs in many VR-based visualization efforts. One early instance was by Fittkau~\cite{Fittkau2015} et al., who implemented a VR experience to aid the PC process that uses the metaphor of a program as a city block: the buildings are classes and packages, and the execution trace is represented as straight-line ``footpaths'' between the buildings. Participants experienced the tool ExplorViz via immersive VR, using gestures to translate, rotate, zoom, and select. The experience was intended to provide analysts a novel tool while employing familiar metaphors. These participants rated the experience of answering basic comprehension questions with this tool as suitable for performing PC, and as an alternative, albeit needing adaptations, to a classic experience.

In another exploration using the city metaphor, Oberhauser and Lecon~\cite{Oberhauser2017} employed immersive VR to aid PC by providing participants the ability to fly through a 3D representation of code in their tool Gamified Virtual Reality FlyThruCode (GVR-FTC). Two metaphors, ``universe'' and ``terrestrial,'' related the code components to familiar concepts, where packages/classes/dependencies were represented as solar systems/planets/light beams and cities/buildings/pipes respectively, and the scale of objects represented various metrics such as number of class methods. The team evaluated the tool using two games that motivated players to comprehend the dependency structure and modularization of a code project compared to a common text editor. Although the subject pool was too small to demonstrate statistical significance, the results did show higher comprehension using VR. 

In a larger-scale experiment, Romano et al.~\cite{Romano2019} compared the relative effectiveness of three tools on the task of source code PC. The baseline was a traditional integrated development environment (IDE) with extensions for code metrics and smells, which was compared to a city-metaphor-based virtual reality environment in both immersive (Code2City\textsubscript{VR}) and non-immersive (Code2City) forms. This implementation of the city metaphor~\cite{Capece2017} creates a building (parallelepiped) for each class, where class properties are reflected in the size and color of the building. In their study~\cite{Romano2019}, 42 participants solved PC tasks based on two large open-source Java projects. Both VR-based tools resulted in significantly better correctness in the completion of PC tasks than the IDE. Additionally, the time to complete the tasks was significantly shorted in the immersive environment than both the non-immersive VR and the IDE. 

More recently, Hoff et al.~\cite{Hoff2022} performed a similar experiment for source code PC comparing their approach, Immersive Software Archaeology, with another VR method and an IDE. Their approach is focused on providing an overview, with multiple levels of abstraction, of a software system's architecture. The higher levels of abstraction (architectural) were represented by solar system/planets/continents, and the lower levels (design) were represented by cities/building/floors. Their study of 54 participants demonstrated that their solution provided similar or better performance in tasks exercising accessing information and finding horizontal and vertical relationships in the system's architecture. 

Finally, we mention a variation of the city metaphor. Weninger et al.~\cite{Weninger2020} introduced the concept of \emph{Memory Cities} to visualize how an application uses heap memory over time, rather than using it to visualize code. Objects on the heap are grouped and represented as buildings in a 3D visualization. Attributes such as color, opacity, area, and height represent various metrics of the heap, and the buildings evolve as the program executes. The authors describe how their tool helps users identify memory leaks in two use cases and plan user studies in the future. 

As mentioned at the start of the section, we are careful to distinguish research contributions that focus on source-level PC from those that focus on binary PC. We include them because, even though we will not have higher-level structures, abstractions, and internal documentation for a binary program, and even though any higher-level abstractions generated by tools will have significant uncertainty (see Section~\ref{section:pchard}), we believe many of the findings presented in this subsection can inform immersive VR for binary PC.

\subsection{Section Summary}

In approaching RQ3, ``How can the affordances of immersive interaction with virtual reality be employed to improve cognition in analytic tasks,'' we reviewed relevant research contributions in the fields of immersive technologies and IA in particular, as well as visualizations and metaphors employed in PC. Through that process we identified many affordances, techniques, and concepts as follows: 
\begin{itemize}
    \item Signalling~\cite{Albus2021}~\cite{Mayer2005}
    \item Incorporate common reverse engineering tools
    \item User-organization of visualizations in 3D space~\cite{Batch2020}
    \item Spatial semantics: spatial organization provides added semantic layer~\cite{Andrews2010}
    \item Incremental formalism: structure is emergent with understanding~\cite{Andrews2010}
    \item Physical navigation: enables efficient access to information through quick body movements~\cite{Andrews2010}
    \item That immersion must be accompanied by embodiment~\cite{Gracanin2018}
    \item Embodied assistant~\cite{deMelo2020}
    \item Context: physical location helps to restore state-of-mind~\cite{Andrews2010}
    \item Persistence: exploits spatial (position and representation) memory to remember information~\cite{Andrews2010}
    \item Use abstractions of environments to which human perception is attuned~\cite{Moloney2018}
    \item Use representations tuned to intermediate zone where human perception is most discerning~\cite{Moloney2018}
    \item Refresh: serendipitous glances refresh memory of information~\cite{Andrews2010}
    \item Awareness: scanning the space quickly assesses overall status~\cite{Andrews2010}
    \item Use a relevant metaphor to visualize programs (e.g., city block)~\cite{Fittkau2015}~\cite{Oberhauser2017}~\cite{Capece2017}~\cite{Averbukh2019}~\cite{Romano2019}~\cite{Hoff2022}
    \item Gamified VR for PC~\cite{Oberhauser2017}
    \item EEG-based adjustment of cognitive load~\cite{Billinghurst2019}
    \item Mimetic references overlaid with constructed affordances~\cite{Moloney2018}
    \item Presence of detail: detailed information enables rapid access and synthesis based on rich content ~\cite{Andrews2010}
    \item Cross-modal mapping~\cite{Moloney2018}
    \item Visualize binary and data files as bitmaps, complementing a text viewer~\cite{Conti2008}
    \item Timeline plots and icons arranged in a spiral, complementing log files~\cite{Gregio2012}
    \item Application of Chunking: collecting lower-level details in a single higher-level abstraction that can help understanding of a large problem~\cite{Waguespack1989}
\end{itemize}

Next, we will analyze our findings regarding VR affordances in the context of our prior findings in cognitive models of binary PC and the cognitive theory behind those models.

\section{Analysis and Next Steps}
\label{section:analysis}

In the framework of CSE, Section~\ref{section:pchard} characterizes the problems of binary PC, setting a context for this work; Sections~\ref{section:cogmodel} and ~\ref{section:cognition} fill the roles of a functional decomposition and work domain analysis; and Section~\ref{section:immersion} describes immersive affordances that may improve human performance in certain applications. Our fourth research question asks, ``How can we use these findings to effect improvements in the practice of binary PC?'' To answer this question, we examine the connections between the findings in the previous sections and identify the most salient themes---performing an indirect cognitive task analysis---in order to propose directions in the information and interaction design for this problem domain.

\subsection{Overview of Elements and Connecting Threads}

\begin{figure*}[htbp]
    \centerline{\includegraphics[width=\textwidth]{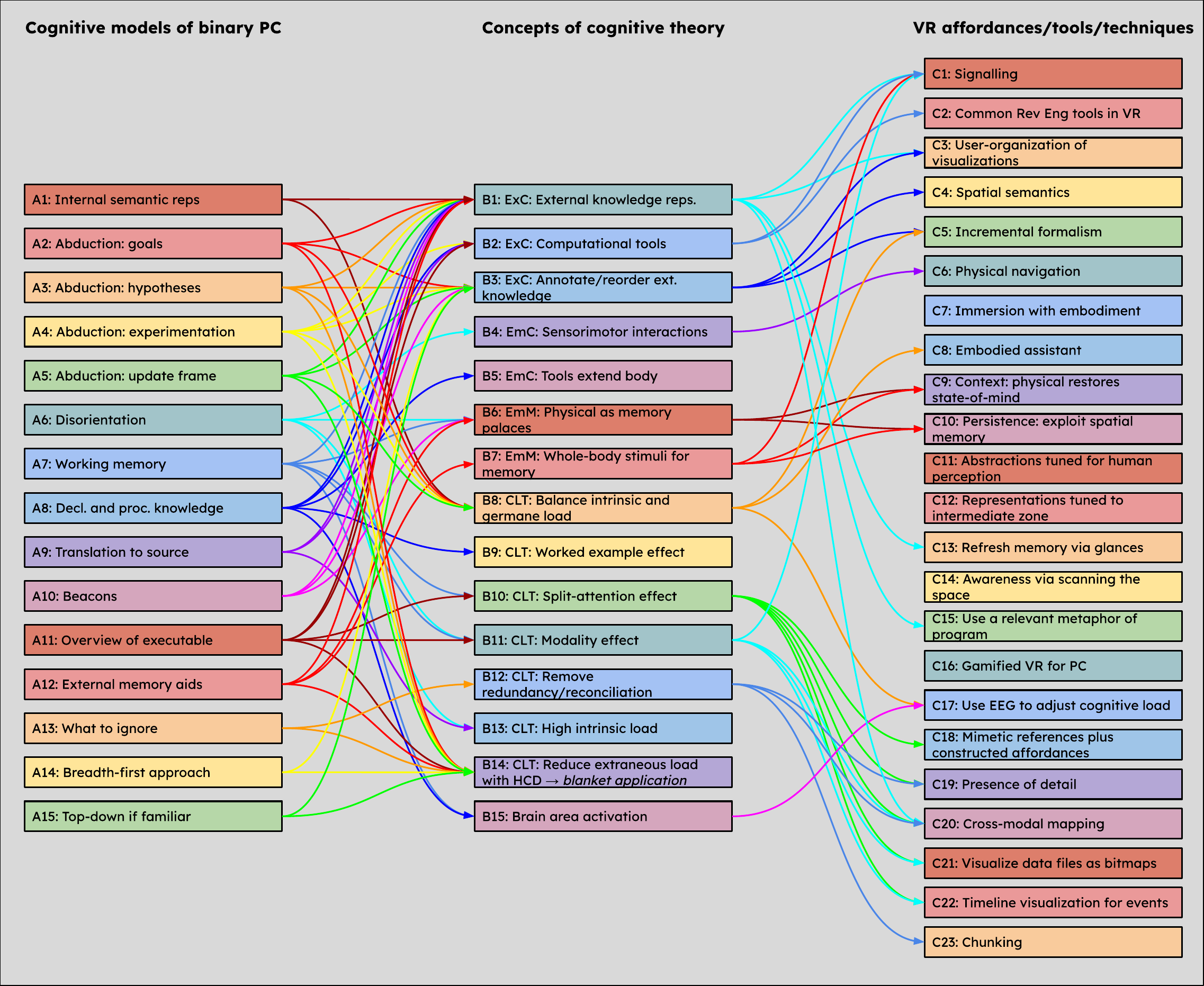}}
    % \centerline{\includesvg[width=\textwidth]{images/Threads.svg}}    
    \caption{Threads connecting elements of (A) cognitive models of binary PC, (B) concepts of cognitive theory, and (C) VR affordances/tools/techniques}
    \label{fig:threads}
\end{figure*}

Consider the findings of this survey partitioned into three groups matching the three previous sections: (A) cognitive models of binary PC, (B) concepts of cognitive theory, and (C) VR affordances/tools/techniques. List~\ref{list:elements} provides the full text of each element with attribution, by group. Conceptual \emph{threads} tie an individual cognitive model element from group A to an element of cognitive theory in group B that supports it, and then to a VR technique in group C that, when implemented, could improve that element of the cognitive model in an immersive binary PC tool. Figure~\ref{fig:threads} provides an overview of these threads tying together the elements from each group (element labels are abbreviated). The authors determined these threads subjectively, linking elements between groups that the authors believe to be sufficiently directly related. We acknowledge that the threshold of ``directness'' is arbitrary and tuned to produce tight groups, and that strong arguments can be made to link the elements in other ways. We present the entire body of threads in the figure to illustrate just one take on the many logical avenues of further research this problem domain provides before we further narrow the set. Next, we will narrow our focus. 

% NOTE: The list below *was* automatically generated from info in a spreadsheet. 
% See "Sankey.ipynb" in "Research\dev\sandbox\sankey and chord chart"
% At this time, just edit it manually as needed. 
\begin{Table*}[h]
\setlength{\columnseprule}{1pt}
\def\columnseprulecolor{\color{black}}
\begingroup
\footnotesize
\begin{multicols*}{3}
\underline{\textbf{Cognitive models of binary PC}}
\begin{enumerate}[label=A\arabic*.,leftmargin=*]
    \item Uses multi-level internal semantic representation of the program~\cite{Shneiderman1979}
    \item Abductive iteration: Sets goals and follows plans~\cite{Bryant2012}~\cite{Nyre-Yu2022} 
    \item Abductive iteration: Forms increasingly complete hypotheses~\cite{Brooks1983}~\cite{Weigand2012}~\cite{Dudenhofer2019}~\cite{Votipka2020}
    \item Abductive iteration: Tests hypothesis through experimentation~\cite{Bryant2012} ~\cite{Sisco2017}~\cite{Dudenhofer2019}~\cite{Votipka2020}
    \item Abductive iteration: Updates framing of the problem based on results~\cite{Klein2007}~\cite{Bryant2012}~\cite{Dudenhofer2019}~\cite{Votipka2020}
    \item Creates disorientation following recursions and execution paths~\cite{Zayour2000}
    \item Taxes working memory~\cite{Shneiderman1979}~\cite{Zayour2000}
    \item Requires declarative and procedural knowledge retrieval and generation~\cite{Bryant2012}  
    \item Uses translation--determining how the code would be implemented in a higher-level language~\cite{Sisco2017}
    \item Uses beacons; beacons for binary PC are more diverse than for source-code-based PC~\cite{Brooks1983}~\cite{Dudenhofer2019}~\cite{Votipka2020}
    \item Relies upon overview of the binary executable~\cite{Votipka2020}
    \item Uses external memory aids~\cite{Detienne2001}~\cite{Storey2005}
    \item Relies upon determining what to ignore~\cite{Mantovani2022}
    \item Experts use a breadth-first approach with system thinking; novices, both depth- and breadth-first without system thinking~\cite{Vessey1985}
    \item Experts use a top-down approach if the program domain is familiar, otherwise, bottom-up~\cite{Siegmund2014}
\end{enumerate}
\columnbreak
\underline{\textbf{Concepts of cognitive theory}}
\begin{enumerate}[label=B\arabic*.,leftmargin=*]
    \item External Cognition: Use external knowledge representations to reduce memory load, e.g., notes and reminders~\cite{Scaife1996}~\cite{Preece2019}
    \item External Cognition: Use computational tools (e.g., calculators) to make tasks easier~\cite{Scaife1996}~\cite{Preece2019}
    \item External Cognition: Annotate and reorder or restructure external representations of knowledge~\cite{Scaife1996}~\cite{Preece2019}
    \item Embodied Cognition: Cognitive processing is influenced by the body and sensorimotor interactions~\cite{Glenberg2013}~\cite{Hornecker2017}~\cite{Shapiro2021}
    \item Embodied Cognition: Tools extend the body schema~\cite{Kirsh2013}
    \item Embodied Memory: Physical objects or locations serve as memory palaces~\cite{Ale2022}
    \item Embodied Memory: Whole-body stimuli can expedite storage and retrieval of memory~\cite{Ale2022}
    \item CLT: Balance immediate problem-solving (intrinsic load) and long-term schema development (germane load)~\cite{Sweller2019}
    \item CLT: Worked example effect of learning from studying solved sample problems~\cite{Hollender2010}
    \item CLT: Split-attention effect of presenting information from multiple sources in an integrated way to reduce load from mental integration~\cite{Hollender2010}
    \item CLT: Modality effect of presenting multiple information sources through different modalities (primarily visual and aural) to reduce the integration load~\cite{Hollender2010}
    \item CLT: Remove redundancy of information presented in different modalities/sources to reduce the load of reconciling the underlying concepts~\cite{Hollender2010}
    \item CLT: Both underloading and overloading can degrade performance; underloading is unlikely due to the high instrinsic load of PC~\cite{Paas2004}~\cite{Helgesson2021}
    \item CLT: Reduce extraneous load: Use human-centered design to reduce the wasted time and friction of poor tools and interactions used to solve a problem~\cite{Helgesson2021}
    \item PC activates areas of the brain associated with working memory, written language, and integration~\cite{Siegmund2014}
\end{enumerate}
\columnbreak
\underline{\textbf{VR affordances/tools/techniques}}
\begin{enumerate}[label=C\arabic*.,leftmargin=*]
    \item Signalling~\cite{Albus2021}~\cite{Mayer2005}
    \item Common Rev Eng tools
    \item User-organization of visualizations in 3D space~\cite{Batch2020}
    \item Spatial semantics: spatial organization provides added semantic layer~\cite{Andrews2010}
    \item Incremental formalism: structure is emergent with understanding~\cite{Andrews2010}
    \item Physical navigation: enables efficient access to information through quick body movements~\cite{Andrews2010}
    \item That immersion must be accompanied by embodiment~\cite{Gracanin2018}
    \item Embodied assistant~\cite{deMelo2020}
    \item Context: physical location helps to restore state-of-mind~\cite{Andrews2010}
    \item Persistence: exploits spatial (position and representation) memory to remember information~\cite{Andrews2010}
    \item Use abstractions of environments to which human perception is attuned~\cite{Moloney2018}
    \item Use representations tuned to intermediate zone where human perception is most discerning~\cite{Moloney2018}
    \item Refresh: serendipitous glances refresh memory of information~\cite{Andrews2010}
    \item Awareness: scanning the space quickly assess overall status~\cite{Andrews2010}
    \item Use a relevant metaphor to visualize programs (e.g., city block)~\cite{Fittkau2015}~\cite{Oberhauser2017}~\cite{Capece2017}~\cite{Averbukh2019}~\cite{Romano2019}~\cite{Hoff2022}
    \item Gamified VR for PC~\cite{Oberhauser2017}
    \item EEG-based adjustment of cognitive load~\cite{Billinghurst2019}
    \item Mimetic references overlaid with constructed affordances~\cite{Moloney2018}
    \item Presence of detail: detailed information enables rapid access and synthesis based on rich content ~\cite{Andrews2010}
    \item Cross-modal mapping~\cite{Moloney2018}
    \item Visualize binary and data files as bitmaps, complementing a text viewer~\cite{Conti2008}
    \item Timeline plots and icons arranged in a spiral, complementing log files~\cite{Gregio2012}
    \item Application of Chunking: collecting lower-level details in a single higher-level abstraction that can help understanding of a large problem~\cite{Waguespack1989}
\end{enumerate}
\end{multicols*}
\endgroup
\caption{Elements of (A) cognitive models of binary PC, (B) concepts of cognitive theory, and (C) VR affordances/tools/techniques}
\label{list:elements}
\end{Table*}

\subsection{Primary Themes}

\begin{figure*}[htbp]
    \centerline{\includegraphics[width=\textwidth]{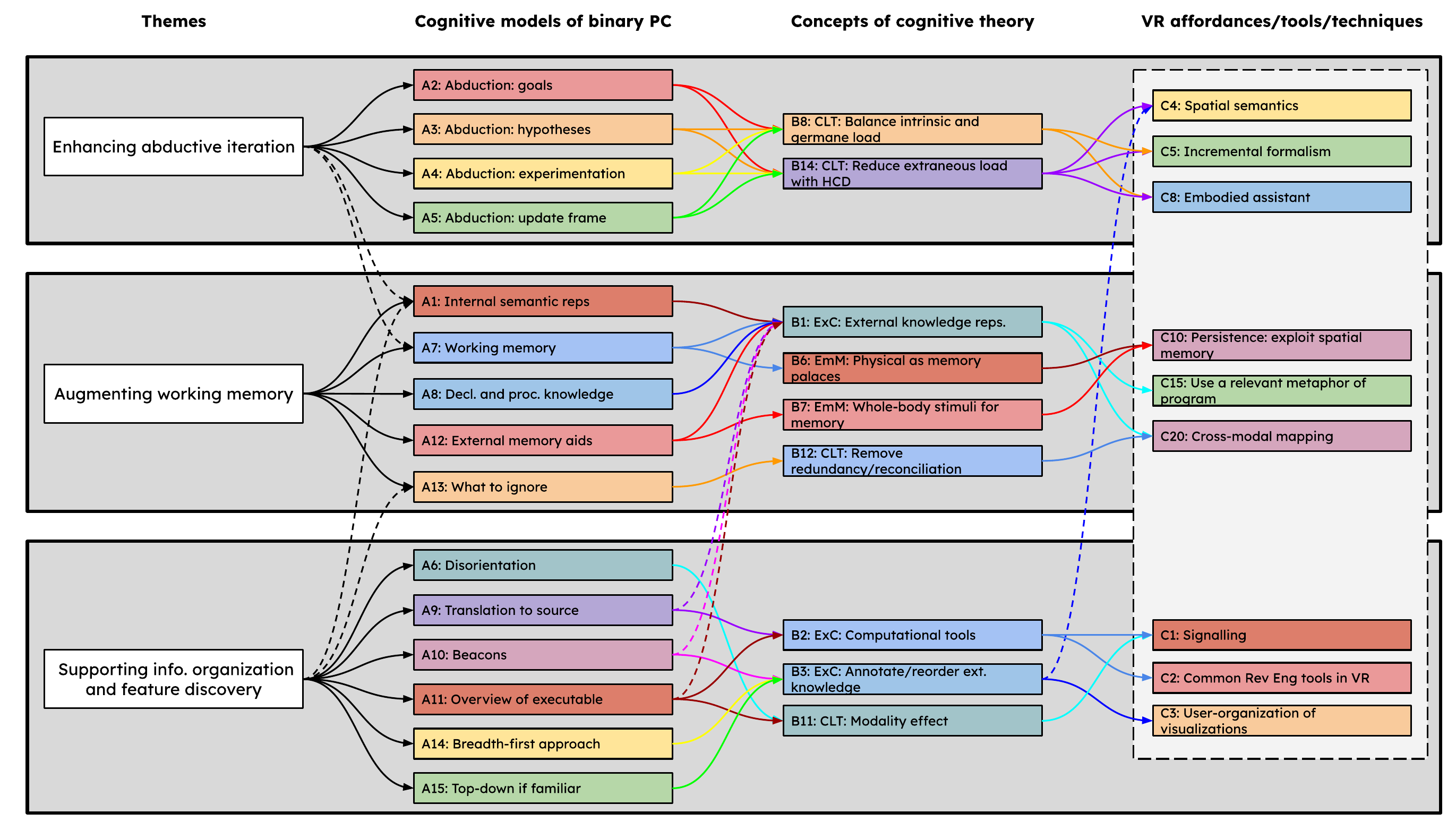}}
    % \centerline{\includesvg[width=\textwidth]{images/Threads_into_Themes.svg}}
    \caption{Primary themes for analysis with most closely-related elements; highlighted area indicates highest-priority VR affordances}
    \label{fig:themes}
\end{figure*}

We aligned related threads into overarching \emph{themes} that are critical to the process of binary PC. In this alignment process, we concentrated on that we expect will have a significant impact on the task of \emph{binary} PC; due to the lack of related research specifically in embodied immersion for binary PC, many of our sources focus on source code PC or software engineering in general, and while we believe many of those concepts may apply in the binary domain, here we've exercised subjective judgement to narrow the focus to those elements that most strongly apply. Clustering these threads revealed the themes we present here:
\begin{itemize}
    \item Enhancing abductive iteration (hypothesis loop)
    \item Augmenting working memory
    \item Supporting information organization and discovery of important features
\end{itemize}
Figure~\ref{fig:themes} depicts the three themes and the most closely related elements from groups A, B, and C. These three themes touch upon every element from group A, cognitive modelling of binary PC, which is important because we want to find the most effective ways to augment as many of the cognitive model elements as possible. Moving into the group B elements of cognitive theory, we are more selective in our areas of concentration. Finally, we derive a central set of VR affordances from group C on which we will focus our efforts in future work. We will examine each theme more closely in the remainder of this section.

\subsubsection{Enhancing abductive iteration (hypothesis loop)}
% Theme: hypotheses / abductive iteration (A2, A3, A4, A5, A1, A7, A12, A13)

One very prevalent theme in cognitive models of PC is the iterative pattern of sensemaking using abductive reasoning. In this analysis, that pattern is broken into four elements of abductive iteration: \emph{setting goals/following plans, forming hypotheses, experimenting to test hypotheses, and updating what is known}~\cite{Bryant2012}~\cite{Nyre-Yu2022}~\cite{Brooks1983}~\cite{Weigand2012}~\cite{Dudenhofer2019}~\cite{Votipka2020}~\cite{Sisco2017}~\cite{Klein2007}. The cognitive model elements of \emph{multi-level internal semantic representation}~\cite{Shneiderman1979} and \emph{heavy dependence on working memory}~\cite{Shneiderman1979}~\cite{Zayour2000} also broadly apply to this theme. 

The elements of cognitive theory most closely related to this theme include managing cognitive load in the iterative sensemaking loop through \emph{balancing the intrinsic and germane loads}~\cite{Sweller2019} and by \emph{reducing extraneous load as much as possible through human-centered design}~\cite{Helgesson2021}. As this theme encompasses the overarching execution of the PC task, it has overlap with the other themes. For example, relevant elements from cognitive theory include those related to information organization and memory. In this theme, we narrow the focus specifically to facilitating the iterative loop and reserve examination of specific elements related to information and memory for later in this section.

In the iterative process, the prime opportunity for augmentation is in amplifying the expert's ability to identify and track progress toward goals and hypotheses. \emph{What is the current goal? What is the current hypothesis? What evidence has been built? What are prior decision points to which we can return upon hitting a dead end?}

The immersive affordances most applicable include \emph{incremental formalism}~\cite{Andrews2010}, in which the visualization of the iterative process evolves its structure as progress is made; \emph{spatial semantics}~\cite{Andrews2010}, in which the spatial organization of the information provides semantic information; and the \emph{embodied assistant}~\cite{deMelo2020} that would (ideally) track the expert's decision path and provide the interactive interface by which the expert can ask questions, make notes, etc. 

\subsubsection{Augmenting working memory}
% Theme: augment working memory (A7, A8, A12)

The limitations of working memory impact the effectiveness and efficiency with which analytical problems are solved, binary PC or otherwise. Closely-related elements from cognitive models of PC include \emph{multi-level internal semantic representation}~\cite{Shneiderman1979}, \emph{taxing working memory}~\cite{Shneiderman1979}~\cite{Zayour2000}, \emph{generation, storage, and retrieval of declarative and procedural knowledge}~\cite{Bryant2012}, \emph{using external memory aids}~\cite{Detienne2001}~\cite{Storey2005}, and \emph{determining what to ignore}~\cite{Mantovani2022}.

The most closely-related elements of cognitive theory include \emph{using external knowledge representations to reduce memory load}~\cite{Scaife1996}~\cite{Preece2019} by offloading some knowledge that otherwise would be held in working memory, \emph{employing memory palaces aided by physical objects or locations and leveraging whole-body stimuli to expedite storage and retrieval of memory}~\cite{Ale2022}, and \emph{removing redundancy of information presented in different modalities}~\cite{Hollender2010} to streamline the intake of new information into working memory.

Following those threads from the cognitive model to cognitive theory leads us to a few considerations for immersive VR: \emph{employing persistence to exploit spatial memory}~\cite{Andrews2010} to remember information, \emph{using physical metaphors}~\cite{Fittkau2015}~\cite{Oberhauser2017}~\cite{Capece2017}~\cite{Averbukh2019}~\cite{Romano2019}~\cite{Hoff2022} for the task's information and operations, and \emph{cross-modal mapping}~\cite{Moloney2018} to present information using multiple senses. Additionally, reducing extraneous cognitive load, as described in the previous theme, can ensure working memory capacity is freed for solving the intrinsic problem.

\subsubsection{Supporting information organization and feature discovery}
% Theme: support information organization and avoid disorientation (A1, A6, A11, A14, A15)
% Theme: discovery of important features (A10, A11, A13)

Of the three themes, \emph{supporting information organization and feature discovery} is the most context-dependent; while the themes of iterative abductive process and enhancements to working memory can apply to almost any analytic problem, this theme is most tightly-integrated with the problem domain and its existing methods and tools. 

The elements from cognitive models of binary PC most closely tied to this problem domain include \emph{disorientation following execution paths}~\cite{Zayour2000}, \emph{translating the binary back to source code}~\cite{Sisco2017}, \emph{using or marking beacons}~\cite{Brooks1983}~\cite{Dudenhofer2019}~\cite{Votipka2020}, \emph{referring to an overview of the binary program}~\cite{Votipka2020}, and \emph{approaching the task in a breadth-first and top-down manner, depending on the expert's familiarity}~\cite{Vessey1985}~\cite{Siegmund2014}. Additional related elements, crossovers with the other two themes, include \emph{multi-level internal semantic representations}~\cite{Shneiderman1979} of the composition of the binary program and \emph{determining what to ignore}~\cite{Mantovani2022} from the masses of data in the program and system.

Several cognitive theory elements support this theme. \emph{Annotating and reordering/restructuring external representations of
knowledge}~\cite{Scaife1996}~\cite{Preece2019} addresses how experts take notes and organize process artifacts. \emph{Using computational tools to make tasks easier}~\cite{Scaife1996}~\cite{Preece2019} covers the use of current well-known reverse engineering tools to discover key characteristics of the binary in an automated or semi-automated fashion. The \emph{modality effect}~\cite{Hollender2010} can also be leveraged given multiple sources of information and interpretations of the same binary program, for example, to provide signalling cues. Finally, as a crossover element with the working memory theme, \emph{using external knowledge representations to reduce memory load}~\cite{Scaife1996}~\cite{Preece2019} acknowledges the need to use the external knowledge representations mentioned at the start of this paragraph. 

Several VR affordances apply to this theme. First, \emph{incorporate common reverse engineering tools} in a way that allows the expert to naturally interact with them from within the immersive environment; this task may require a novel interaction paradigm. Second, \emph{signalling}~\cite{Albus2021}~\cite{Mayer2005} will help the expert focus on what is most important; what is signalled may come from self-sourced notations, or it may come from cues inferred from the reverse engineering tools. \emph{Spatial semantics}~\cite{Andrews2010} applies here just as in the iterative abduction theme: the spatial organization of the data provides meaning itself, over and above the meaning of the data. Finally, \emph{user organization of visualizations}~\cite{Batch2020} applies, especially in combination with the preceding element, where that organization in space is itself meaningful.

\subsection{Section Summary}

Three primary themes capture the salient aspects of cognitive models of binary PC. Following the threads that connect cognitive models to cognitive theory and then to VR affordances, we derived a set of the most relevant qualities for an immersive experience for binary PC:
\begin{itemize}
    \item\emph{Incremental formalism}~\cite{Andrews2010}, in which the visualization of the iterative process evolves its structure as progress is made 
    \item\emph{Spatial semantics}~\cite{Andrews2010}, in which the spatial organization of the information provides semantic information
    \item\emph{Embodied assistant}~\cite{deMelo2020} that (ideally) tracks the expert's decision path and provides the interactive interface by which the expert can ask questions, make notes, etc. 
    \item\emph{Employing persistence to exploit spatial memory}~\cite{Andrews2010} to remember information
    \item\emph{Meaningful and relatable metaphors}~\cite{Fittkau2015}~\cite{Oberhauser2017}~\cite{Capece2017}~\cite{Averbukh2019}~\cite{Romano2019}~\cite{Hoff2022} for the task's information and operations
    \item\emph{Cross-modal mapping}~\cite{Moloney2018} to present information using multiple senses
    \item\emph{Incorporate common reverse engineering tools} to leverage powerful tools already familiar to experts
    \item\emph{Signalling}~\cite{Albus2021}~\cite{Mayer2005} focuses the expert's attention on potential key discoveries
    \item\emph{User organization of visualizations}~\cite{Batch2020}, in conjunction with spatial semantics, allows the expert to capture knowledge in the positioning of the data itself 
\end{itemize}

In addition to these specific elements, we must consider overarching considerations for an immersive VR implementation. Any successful implementation will follow an \emph{iterative human-centered design and development process}. It will incorporate automation as much as possible, including taking advantage of large language models. Additionally, it will be \emph{adaptive} and self-tune the parameters of the immersive experience to give the expert user the most effective experience. 

The actual visualizations, environments, interactions, and metaphors to be used in an immersive experience still require investigation; these qualities should serve as guidelines heavily influencing the selection, design, and development of those experiences. Some features to consider implementing following the above guidelines include:
\begin{itemize}
    \item \textbf{Intuitive representation of the code}, highlighting features like function calls and branching/switch statements
    \item \textbf{Basic discovery of the binary}: the compiler that built the binary; clusters of system calls; presence of encryption functions; places where a code block is referenced; etc.
    \item \textbf{Identifying, marking, and tracking beacons}; point, click, annotate to create
    \item A simplified environment, \textbf{masking out what can be ignored}; allowing the user to ``collapse'' and ``expand'' portions of the data
    \item Providing an expedited way to get an \textbf{overview of the complete binary program} and then drilling down
    \item Expediting the \textbf{creation, storage, and retrieval of external knowledge} through a natural interface
\end{itemize}

\section{Conclusion}
\label{section:conclusion}

The process of binary PC is difficult and requires very specialized expertise to perform it effectively. The problem is only getting harder as computer architectures become more varied and complex and binary obfuscation techniques become more sophisticated. Augmenting the cognitive process of binary PC is crucial to maintaining or improving the current level of effectiveness of experts performing this task. 

Following a CSE approach, we surveyed prior work in three progressive groups to support an indirect cognitive task analysis: characteristics of mental or cognitive models of binary PC, cognitive theory and applications to binary PC, and cognitive augmentation using visualization and immersive technologies. We identified several common or important elements of each group and presented threads of related elements. Through an inductive process, we identified three primary themes that encapsulate the elements of cognitive models surveyed, related threads of cognitive theory, and immersive affordances that provide a path for implementing cognitive augmentation of the binary PC process. Finally, we derived a set of guidelines by which to design and develop immersive experiences for binary PC.

\section{Acknowledgments}

The primary author is supported by the United States Air Force Science \& Engineering Career Field Team Science, Technology, Engineering, Mathematics + Management (STEM+M) program.

\section{Author Disclaimer}

The opinions contained herein are those of the authors only and should not be construed as those of the Department of Defense or United States Government.

% -------------------

\bibliographystyle{acm/ACM-Reference-Format}
% -------------------
% BIBLIOGRAPHY IMPORT
\bibliography{VR.bib}
% -------------------

\end{document}